\documentclass[preprint,showpacs,a4paper,superscriptaddress,floatfix]{revtex4-1}
\usepackage{amsmath}
\usepackage{amssymb}
\usepackage{graphicx}

\usepackage[english]{babel}
\usepackage{longtable}
\usepackage{bm}
\usepackage{epstopdf}
\usepackage{epsfig}
\usepackage{subfigure}
\usepackage[caption=false]{subfig}

\begin{document}

\title{Band folding, strain, confinement, and surface relaxation effects on the electronic structure of GaAs and GaP: from bulk to nanowires}

\author{Cl\'audia Lange dos Santos}
\email{claudialange@unifra.br}
\affiliation{\'Area de Ci\^encias Tecnol\'ogicas, Universidade Franciscana, 97010-032, Santa Maria, RS, Brazil and \\ Dipartimento di  Fisica, Informatica e Matematica (FIM) dell'universit\'a degli studi di
  Modena e Reggio Emilia, via Campi
  213/A, 41100 Modena, Italy}

\author{Paulo Piquini}
\affiliation{Departamento de F\'{i}sica, Universidade Federal de Santa Maria, 97105-900, Santa Maria, RS, Brazil}

\author{Rita Magri}
\affiliation{Dipartimento di  Fisica, Informatica e Matematica (FIM)  dell'universit\'a degli studi di
 Modena e Reggio Emilia and  CNR-Nano-S3, via Campi
 213/A, 41125 Modena, Italy}

\begin{abstract}

It is common to find materials that show strikingly different properties between its bulk and nanometric forms. In this paper we show how to link the electronic structures of two III-V systems, one a direct gap material, GaAs, and the other an indirect gap material, GaP, from their bulks right down to the shape of thin nanowires. The understanding of how these changes occur represents a scientific and technological challenge and is relevant for the design and prediction of novel nanostructured materials. GaAs and GaP bulk and nanowire systems are studied in the zinc-blende and wurtzite structures both free of strain and subjected to biaxial strains perpendicular to the [111]/[0001] direction (the kind of strain that the materials are subjected to when grown one on top of the other). We provide an interpretation of the band structure of nanowires, grown along the [111] (zinc-blende structure) and the [0001] (wurtzite structure) directions, in terms of the bulk band structures of the corresponding binary compounds. The procedure reveals the origin of the valence and conduction valleys relevant to determine the nature (direct or indirect) of the band gaps and the kind (direct and pseudodirect) of the valence to conduction transitions. Thus, by calculating only the bulk bands it is possible to describe the behavior of the nanowire bands even for very thin nanowires. The effects on the band structures due to biaxial strain are analogously analyzed, providing for bulk GaP the first results in literature. The role of confinement, and surface relaxation, in determining the nanowire electronic structure of thin nanowires are analyzed separately revealing that the change in the nature of the band gap is due mainly to surface relaxation effects, not confinement. We show that the change for indirect/direct of the gap from the bulk to the 1D systems is mainly due to the competition between the energies of bulk conduction valleys which are differently influenced by confinement and strain. This effect is shown also by other low dimensional materials like the 2D materials extending only few atomic layers in one dimension. The competing valleys are already present in the bulk band structure. While the main effect of confinement is to open all gaps it is not necessarily the main cause of the direct/indirect change in the nature of the electronic gap as instead is usually claimed in the literature. Our study can be used to understand and engineer the structure of many nanostructures systems by just better analysing the behavior of the bulk bands.

\end{abstract}


\date{\today}

\maketitle

%%%%%%%%%%%%%%%%%%%%%%%%%%%%%%%%%%%%%%%%
\section{Introduction}
%%%%%%%%%%%%%%%%%%%%%%%%%%%%%%%%%%%%%%%%
The comprehension of how different materials properties transform when the spatial dimensions are reduced is not an obvious task. Experimentally, it would require a systematic study that monitors the evolution of the properties of several samples over the whole range of sizes and shapes between the bulk and the smallest nanopieces of a given material. Further, is not easy for the experimentalist to separate out the influences of the different factors concurring to produce a given result for each analyzed sample. Some electronic properties, in particular, can only be properly understood through appropriate theoretical treatments, e.g., the nature and ordering of the electronic levels in each sample. Ab initio theoretical approaches allow for a precise description of materials properties for samples at the two extremes, the infinite size bulk system (with translational symmetry) and the nanosamples of a given system. In addition, it is possible for the theorists to isolate the different factors contributing to produce a given measured result. Hence, an appropriate theoretical study can allow for an unique understanding of how the properties of bulk and nano samples of a given material are related, and how these relationships change from material to material.

Semiconductor nanowires (NWs) are considered promising systems for different kinds of technological applications including, for instance, light emitting diodes (LEDs)\cite{Berg}, lasers\cite{Li}, solar cells\cite{Aberg,Yao}, and high electron mobility transistors (HEMTs).\cite{Konar,Shen} One of the advantages in the use of these nanomaterials is related to their synthesis process, which allows a precise control of their characteristics.\cite{Zhang} In addition, the large surface to volume ratio of NWs enables an efficient strain relaxation that makes it possible to grow nanowire heterostructures (NWHs) from materials with mismatched lattice constants, a fact that would not be possible in the conventional 2D films. 
For the III-V NWs, in particular,  the exploitation of the allowable degrees of freedom such as compositions, structural politypes, orientations, diameters, surface passivation, and doping, can be used to tune their electronic structure properties. It is therefore decisive to understand how these various variables influence the electronic structure of a NW, taking as the starting point the usually well known electronic structures of the bulk systems. 

In this work we focus on GaAs and GaP NWs. The reason is that these NWs show a huge potential for nanophotonics and nano-optoelectronic applications. GaAs NWs are lattice matched to Ge while GaP NWs to Si,  allowing for a nice integration with the micro-electronic circuits. GaAs NWs have been extensively studied whereas GaP NWs have been studied much less. GaP NWs have been shown to emit in the green with a very intense photoluminescence signal even if the expected first emission has been shown to be dark.\cite{Greil} 
High quality heterostructured NWs built from GaAs, GaP and GaAsP were grown\cite{Wu}, both in the axial geometry as well as in the core-shell one. Also high quality GaAs NWs with mixed zinc-blende and wurtzite structure have been grown.\cite{Spirkoska}  We are mainly interested on how the band edges around the main gap between occupied and empty electronic states develop starting from the electronic structure of the binary compounds. In a previous study Peng et al. have shown how the type of the band gap of GaAs NWs (direct or indirect) is determined by the competition between  different valleys at the conduction band edge whose energies can be tuned by subjecting the NW to a given strain.\cite{Peng} We investigate the origin and characteristics of the conduction and valence band valleys for both GaAs and GaP NWs. In their bulks GaAs is a direct gap semiconductor, whereas GaP is an indirect gap semiconductor. We analyze the effects on the conduction and valence edge valleys due to the structure (zinc-blende (ZB) or wurtzite ( WZ)), dimensions, confinement, sidewall atomic relaxation, and biaxial strain (the kind of strain arising when the materials are grown one on top of the other). We also compare the electronic structures of  GaAs NWs with those of the much less studied GaP NWs.

Details of the employed methodology are shown in Sec. II. The results for the unstrained binaries and strained binaries are shown in Sec. III A and III B. The electronic structures of the NWs at their minimum energy configurations are presented in Sec. III C. Section III C.1 analyze how the valence and conduction edge valleys of the NW band structures evolve from that of the binary compounds. Section III C.2 focuses on the roles of confinement and surface relaxation, separately, in determining the nature of the NW gap (direct and indirect). Finally, our conclusions are summarized in Sec. IV.

%%%%%%%%%%%%%%%%%%%%%%%%%%%%%%%%%%%%%%%
\section{Methodology}
%%%%%%%%%%%%%%%%%%%%%%%%%%%%%%%%%%%%%%%%

The first principles calculations are based on the Density Functional Theory (DFT)\cite {Hohenberg, Kohn} as implemented in the open source package QUANTUM ESPRESSO (http://www.quantum-espresso.org). \cite{Giannozzi} For the electronic exchange and correlation potential we used the local density approximation. \cite{Ceperley, Perdew} The interaction between the valence electrons and the atomic cores are described by separable norm-conserving core-corrected pseudopotentials. \cite{Kleinman, Hamann} These pseudopotentials have shown to be quite efficient (convergence at a small cutoff and prediction of good structural properties) in a number of occasions.\cite{Magri1, Magri2} The Kohn-Sham (KS) wave functions are expanded in plane waves with a cutoff energy of 40 Ry. The equilibrium geometries are obtained when the atomic forces are smaller than $10^{-7}$ Ry/Bohr and the total energy converges within $10^{-6}$ Ry. k-point Monkhorst-Pack grids \cite{Monkhorst} are used for the Brillouin zone sampling. For GaAs and GaP binary compounds with the ZB (WZ) structures, a mesh of $4\times4\times4$ ($4\times4\times2$) was used.  For the NWs, a $1\times 1\times N$ mesh was used, with $N = 8$ in the case of WZ NWs and $N = 6$ in the case of ZB NWs.

%%%%%%%%%%%%%%%%%%%%%%%%%%%%%%%%%%%%%%%
\section{Results}
%%%%%%%%%%%%%%%%%%%%%%%%%%%%%%%%%%%%%%%%

%%%%%%%%%%%%%%%%%%%%%%%%%%%%%%%%%%%%%%%
\subsection{GaAs and GaP binary systems}
%%%%%%%%%%%%%%%%%%%%%%%%%%%%%%%%%%%%%%%%i

Since the DFT-LDA scheme leads to a general underestimation of the band gaps the aim of this section is to investigate the possibility to obtain meaningful trends for the electronic structure of the GaAs-GaP compound systems. Calculations are carried out for the binary compounds and, wherever possible, comparisons are made with other theoretical works and experimental results.

For the ZB structures, we have calculated the total energy as a function of the lattice parameter using a 40 Ry cutoff on the plane waves expansion, and have fit the results to the Murnaghan equation. \cite{Murnaghan} For the WZ structures, instead,  to obtain the equilibrium geometry (lattice parameters $a$ and $c$ need to be optimized simultaneously), we have calculated the total energy using a cutoff energy of 150 Ry and an optimization procedure for both the cell parameters and the atomic positions. The results for the structural parameters and the band gaps of the bulk materials are shown in Table \ref{zb_unstrained}. The WZ compounds do not exist at normal ambient conditions and their data have been calculated only as a reference to the results obtained on the corresponding NWs (where the dominant phase at relatively small diameters is actually the WZ phase).  

\begin{table}[h!]
\caption{\label{zb_unstrained} Calculated equilibrium lattice constants, and band gaps for GaAs and GaP in ZB and WZ phases. The experimental values were taken from Reference \onlinecite{Madelung}. The labels D and I stands for direct and indirect band gap, respectively.}
\begin{ruledtabular}
\begin{tabular}{ccccccccc}
System & $a$({\AA}) & $a^{\mathrm{expt}}$ (\AA) & $l$ (\AA) & $c$ (\AA) & $c/l$ & $E_{g}$(eV) & $E_{g}^{\mathrm{expt}}$(eV)  \\ \hline
GaAs - ZB & 5.606  & 5.653  & -- & -- & -- & 0.800(D) & 1.519  \\
GaP - ZB & 5.400  & 5.451  & --  & -- & -- & 1.415(I) & 2.350  \\
GaAs - WZ & 5.585 & -- & 3.949 & 6.516 & 1.649966 & 0.801(D) & -- \\
GaP - WZ & 5.380 & -- & 3.805 & 6.276 & 1.649610 & 1.334(D) & --\\
\end{tabular}
\end{ruledtabular}
\end{table}

As we can see in Table \ref{zb_unstrained}, the equilibrium lattice constants for the binaries in the ZB structure are in good agreement with the experimental values. The LDA underestimation of the lattice parameters ($\Delta_a$) is 0.047 {\AA} for GaAs and 0.051 {\AA} for GaP, similar for the two binaries. The underestimation of the band gaps ($\Delta_{E_g}$) is 0.719 eV for GaAs (52.7$\%$ smaller than the experimental value at T $\sim$ 0$^{\circ}$ K) and 0.935 eV for GaP (indirect gap) (60.2$\%$ smaller). These underestimations of the main gaps between the valence and the conduction bands are similar for the two materials.

The calculated value of the $c/l$ ratio is larger than the ideal one $\sqrt{\frac{8}{3}} = 1.6329932$, which suggests that the WZ phase is not the stable polymorph of GaAs and GaP, in agreement with the rule stated by Yeh et al..\cite{Yeh} We notice that the lattice parameters $a = l\sqrt{2}$ are consistently smaller in the WZ structure than in the corresponding ZB structure.  These structural parameters are in agreement with the literature. \cite{Belabbes,Tawinan,Panse} 

\begin{figure}[h!]
\centering
\includegraphics[width=12 cm,clip]{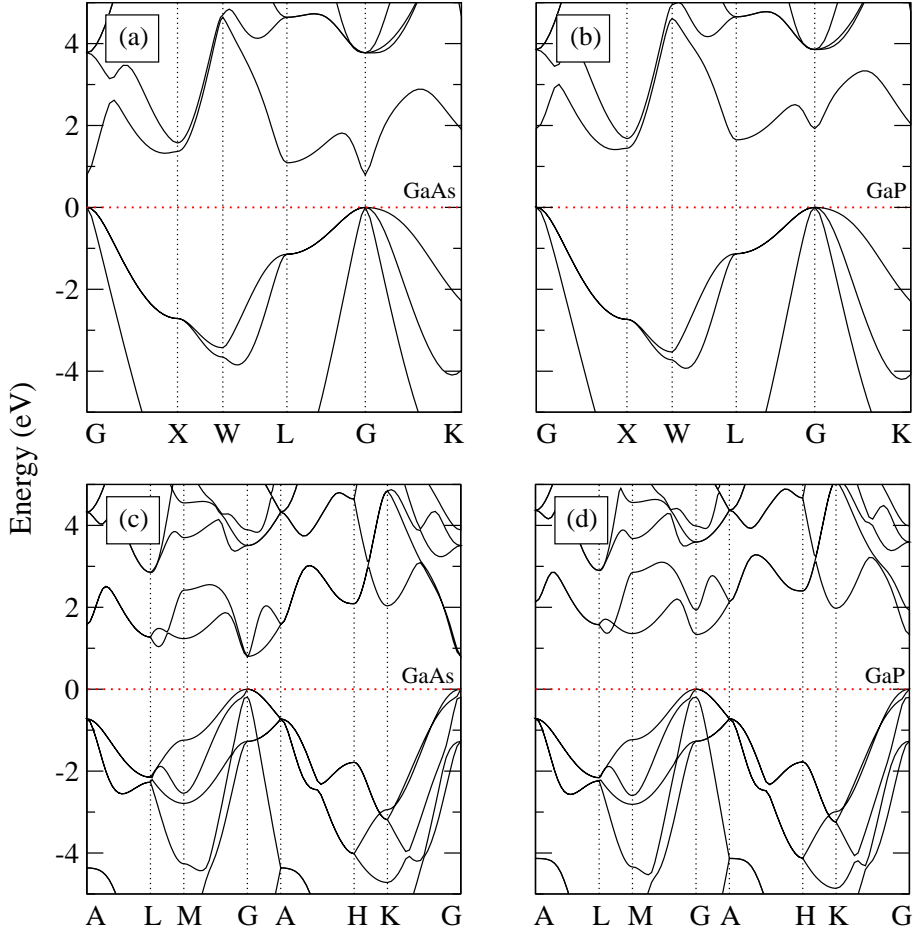}
\caption{(Color online) Calculated band structures for zinc-blende and wurtzite bulk materials: (a) ZB GaAs, (b) ZB GaP, (c) WZ GaAs and (d) WZ GaP.}
\label{bands_zbwz}
\end{figure}

Fig.\ref{bands_zbwz}(a) and  Fig.\ref{bands_zbwz}(b) show, respectively, the electronic band structures of the GaAs and GaP ZB binary systems at their equilibrium atomic configuration while Fig.\ref{bands_zbwz}(c) and  Fig.\ref{bands_zbwz}(d) show the band structures of the same two binary systems along the symmetry directions of the hexagonal Brillouin Zone of the WZ structure.
 
In the ZB phase, GaAs is a direct gap material while GaP is an indirect gap semiconductor with the conduction band minimum (CBM) near the X-point (at $k=0.85 \quad (2\pi/a)$ along the $\Delta$ direction) as shown in Fig. \ref{bands_zbwz}(b). The calculated band gap for GaP at the G-point is 1.935 eV, the band gap difference between this value and the minimum gap is 0.52 eV. The experimental band gap at the G-point is 2.895 eV, with the experimental difference between direct and indirect gap being 0.545 eV, close to our theoretical result. 

In the WZ phase, the band gap value is similar to that of the ZB phase in the case of GaAs, while for GaP it becomes direct and smaller. Similar trends were found by Belabbes et al. \cite{Belabbes} with the band gaps calculated using the LDA-1/2 method. \cite{Ferreira1,Ferreira2} Actually, Yeh, Wei and Zunger \cite{zunger} established three rules to predict the band structure of a WZ compound from its ZB energy levels. For GaAs, the authors found that the corresponding WZ system will be direct with a slightly larger gap. For GaP, on the other hand, the band gap becomes pseudodirect in the WZ phase. This indirect-pseudodirect transition occurs because the L$_{1C}$ state is close to the CBM at X$_{1C}$. A direct experimental comparison between ZB and WZ bulks is not possible since WZ bulks are not available. There is no a definitive  conclusion in the literature about the values obtained for the band gaps of GaAs in the WZ and ZB phases. \cite{Tawinan,De,Heiss,Ketterer2,Regolin,Soshnikov,Vainorius,Zanolli,Kusch,Kusch2,Martelli,Hoang,Jahn,Hjort,Ahtapodov,Gurwitz}

In Table \ref{zb_comparison} we report the calculated and experimental energy band gaps at different k-points. For GaAs, we have found differences between these two quantities of  0.72, 0.61 and 0.73 eV at the G, X and L points. For GaP, these quantities are 0.97, 0.90 and 1.07 eV. In percentage terms, the obtained values for these differences were 47$\%$ (34$\%$) at the G-point, 31$\%$ (38$\%$) at the X-point and 40$\%$ (39$\%$) at the L-point for GaAs (GaP). The underestimation of the gaps is seeing to have almost the same percentual magnitude for GaAs and GaP over all the Brillouin zone.
 
\begin{table}[!h]
\caption{\label{zb_comparison} Calculated band gaps (in eV) at G, X and L points  for zinc-blende GaAs and GaP binaries compared with experimental values taken from Reference. \onlinecite{Vurgaftman} }
\begin{ruledtabular}
\begin{tabular}{ccccc}
System & Method & \multicolumn{3}{c}{$E_g$ (eV)} \\ \cline{3-5}
       &        &      G  & X  & L  \\

GaAs & Present work  & 0.80 & 1.37 & 1.09  \\
     & Experiment    & 1.52 & 1.98 & 1.82   \\
GaP  & Present work  & 1.89 & 1.45 & 1.65  \\ 
     & Experiment    & 2.86 & 2.35 & 2.72
\end{tabular}
\end{ruledtabular}
\end{table}

These results for the binary compounds reassure us that despite the gap underestimation the main trends in the electronic structures can be realistically appreciated in our LDA calculations.

%%%%%%%%%%%%%%%%%%%%%%%%%%%%%%%%%%%%%%%
\subsection{Biaxially strained GaAs and GaP binary systems}
%%%%%%%%%%%%%%%%%%%%%%%%%%%%%%%%%%%%%%%%

When few layers of a material A are grown on top of a material B whose lattice constants do not precisely match, the accumulated stress can be relaxed by forming structural defects that alter the electronic and optical properties of the combined system. NWs, because of the lateral surfaces, can better adjust the stress due to the lattice misfit between the two materials, through lateral strains at the NW surface, and the top layers can grow in a pseudomorphic way without forming such defects at the interface. Thus, some combinations of materials can be obtained by growing one on top of the other in the NW form, but not in the bulk.

Since III-V NWs grow along the [111] (ZB phase) and [0001] (WZ phase) directions, we consider the biaxial stresses on the planes perpendicular to these directions. For the case of GaP, in particular in the WZ structure, these are the first results in the literature.

For this study we will be considering hexagonal unit cells for both the ZB and the WZ structures. 
The total energy of the binaries is minimized as a function of the lattice parameter along the [111] and [0001] directions keeping the in-plane lattice constant $a_{\parallel}$ fixed. The calculations are repeated for different $a_{\parallel}$, where both, tensile and compressive strains have been considered.
 
\begin{figure}[h!]
\centering
\includegraphics[width=16cm,clip]{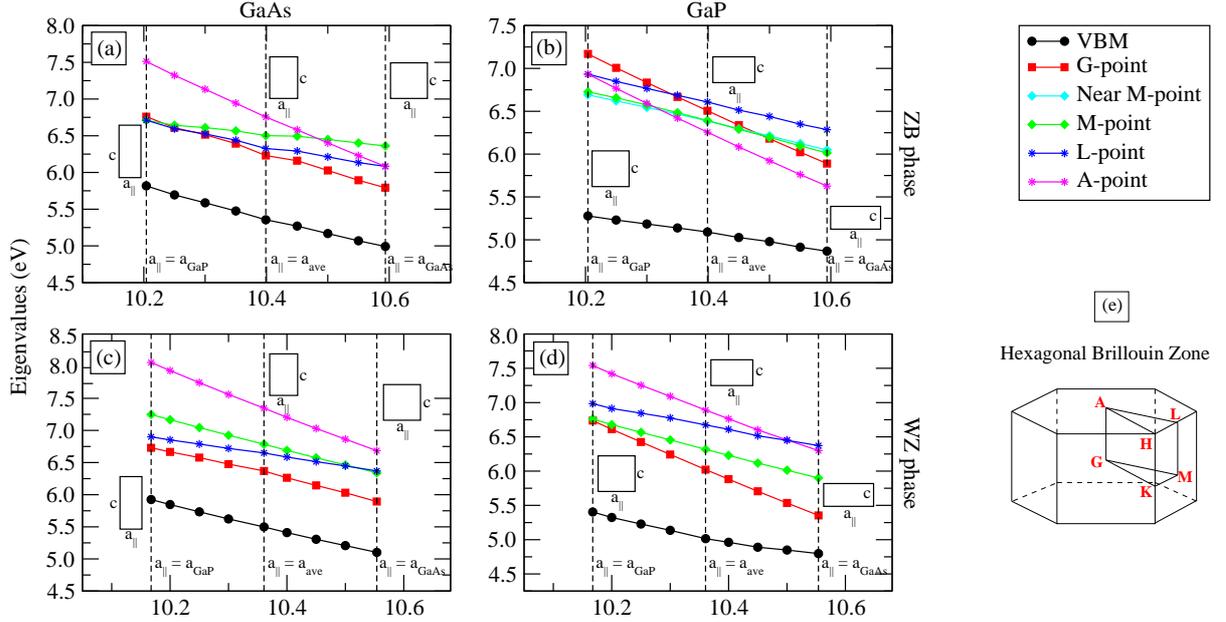}
\caption{(Color online) Eigenvalues associated to the VBM at the G point and to the lowest unoccupied states in the conduction band at different high symmetry k-points, as a function of different values of the in-plane lattice parameter for the GaAs and GaP binaries in the ZB and WZ structures. (a) ZB GaAs, (b) ZB GaP, (c) WZ GaAs and (d) WZ GaP. The irreducible Brillouin zone of the hexagonal cell is shown  in part (e) of the figure.}
\label{levels}
\end{figure}

The results are given in Figure \ref{levels}, which shows the band edge energies of the GaAs and GaP systems at different k points of their Brillouin zone, as a function of the in-plane lattice constant $a_{\parallel}$. 

For the case of the WZ structure there is not change of the crystal symmetry, which remains hexagonal under biaxial stress in the plane orthogonal to the [0001] direction. For the ZB structure, on the other hand, the stress reduces the crystal symmetry from the cubic symmetry to the hexagonal one. This leads to a change of the Brillouin Zone of the ZB structure from the truncated octahedron to the hexagonal one. The corresponding high symmetry points G, X, L of the ZB Brillouin Zone are remapped into k-points of the hexagonal Brillouin Zone, as shown in Figure \ref{levels}. The X points of the ZB Brillouin Zone go onto the M points of the hexagonal Brillouin Zone. In the same way, the two L points, with coordinates (1/2, 1/2, 1/2) and (-1/2, -1/2, -1/2) along the direction of folding, become the A points, while  the other six L points are still labeled L points in the hexagonal Brillouin Zone. 

We can see from Fig. \ref{levels} that in general the compression of the in-plane lattice parameters leads to a shift of the eigenvalues towards higher energies, while a dilation leads to a lowering of the eigenvalues. Furthermore, we notice that the trends are almost linear in all cases, albeit with a different slope for each level. Fig. \ref{levels} (a) shows that ZB GaAs switches from a direct to an indirect gap when its in-plane lattice parameter squeezes circa 3.2$\%$. The conduction band minimum changes from the G to the L ({\bf A}) (direct to indirect band-gap transition) and then from L ({\bf A}) to X ({\bf M}). On the other hand, ZB GaP remains an indirect gap material for all $a_{\parallel}$ values between the lattice parameters of GaP and GaAs, with CBM changing from near the X ({\bf M}) point to the L ({\bf A}) point when the GaP in-plane lattice constant stretches by 1.4 $\%$. All these effects occurs due do the different responses from the energy levels to biaxial strains (different biaxial deformation potentials)

On the contrary the binaries in the WZ structure Fig. \ref{levels} (c),(d) remain always direct gap materials. The only relevant difference between GaAs and GaP is the steeper slope of the conduction minimum at G in GaP than in GaAs (a feature presented also by the ZB structure). Also, in all cases the A point state is the most sensitive to the biaxial strain.

In Table \ref{zbwz_strained} we report the $c$ lattice parameters along the [111] and [0001] directions and the band gaps of the binary GaAs and GaP biaxially constrained to match two in-plane lattice constants: (i) the lattice constant of GaP bulk ($a_{\parallel} = a_{\scriptsize{\mbox{GaP}}}$), and (ii) the average lattice constant ($a_{\parallel} = a_{\scriptsize{\mbox{ave}}}$) where $a_{\scriptsize{\mbox{ave}}} = (a_{\scriptsize{\mbox{GaAs}}}+ a_{\scriptsize{\mbox{GaP}}})/2$. For the ZB structures, $a_{\scriptsize{\mbox{GaP}}}$ = 5.400 {\AA} and $a_{\scriptsize{\mbox{ave}}}$ = 5.503 {\AA}. For the WZ structures, $a_{\scriptsize{\mbox{GaP}}}$ = 5.380 {\AA} and $a_{\scriptsize{\mbox{ave}}}$ = 5.483 {\AA}.

\begin{table}[h!]
\caption{\label{zbwz_strained} Equilibrium lattice constant and band gaps for the strained binaries in the ZB and WZ phases. The labels D and I stands for direct and indirect band gap, respectively.}
\begin{ruledtabular}
\begin{tabular}{ccccccccc}
& \multicolumn{4}{c}{ZINC-BLENDE} & \multicolumn{4}{c}{WURTZITE} \\
  & \multicolumn{2}{c}{$a_{\parallel} = a_{\scriptsize{\mbox{GaP}}}$} & \multicolumn{2}{c}{$a_{\parallel} = a_{\scriptsize{\mbox{ave}}}$} & \multicolumn{2}{c}{$a_{\parallel} = a_{\scriptsize{\mbox{GaP}}}$} & \multicolumn{2}{c}{$a_{\parallel} = a_{\scriptsize{\mbox{ave}}}$} \\ \cline{2-3} \cline{4-5} \cline{6-7} \cline{8-9}
System    &  $c$({\AA})  & $E_{g}$ (eV) & $c$ ({\AA}) & $E_{g}$(eV) & $c$({\AA})  & $E_{g}$ (eV) & $c$ ({\AA}) & $E_{g}$(eV) \\  \hline
GaAs & 9.878  & 0.879(I) & 9.827 & 0.879(D) & 6.622 & 0.695(I) & 6.569 & 0.872(D) \\
GaP  & 9.360  & 1.457(I) & 9.276 & 1.166(I) & 6.276 & 1.334(D) & 6.235 & 1.016(D) \\
\end{tabular}
\end{ruledtabular}
\end{table}

In the case $a_{\parallel} = a_{\scriptsize{\mbox{GaP}}}$, GaAs is under an in-plane compressive strain so it stretches along the growth direction whereas GaP is unstrained.  For $a_{\parallel} = a_{\scriptsize{\mbox{ave}}}$, GaAs is under a lower in-plane compressive strain while GaP is under an in-plane tensile strain so its lattice parameter shortens along the [111] and [0001] directions.

The band gaps of GaAs in the ZB and WZ phases are both direct at $a_{\parallel} = a_{\scriptsize{\mbox{ave}}}$ and indirect at $a_{\parallel} = a_{\scriptsize{\mbox{GaP}}}$. For the GaP, on the other hand, the band gaps are indirect at both values of $a_{\parallel}$ in the ZB phase and direct in the WZ phase.

%%%%%%%%%%%%%%%%%%%%%%%%%%%%%%%%%%%%%%%
\subsection{ GaAs and GaP Nanowires}
%%%%%%%%%%%%%%%%%%%%%%%%%%%%%%%%%%%%%%%%

\begin{figure}[h!]
\centering
\includegraphics[width= 12 cm,clip] {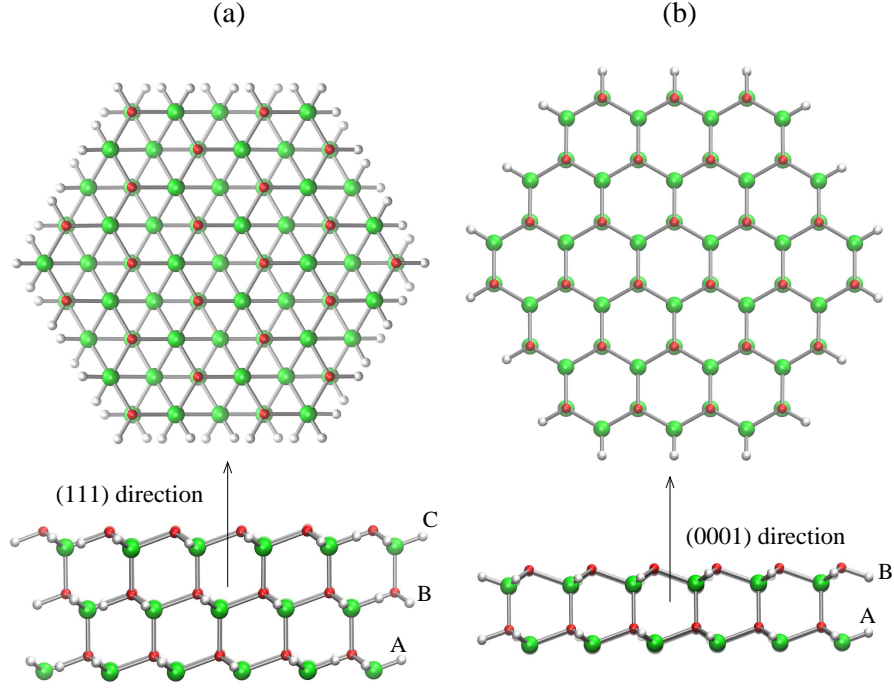}
\caption{(Color online) Representative cross sections and lateral views of a (a) ZB and (b) WZ NW grown along the [111] and [0001] directions, respectively. The larger green (gray) balls represent the As (P) atoms, the medium red balls (dark gray) represent the Ga atoms and the smaller white balls represent the pseudohydrogen atoms.}
\label{nanowires}
\end{figure}

We will now analyze GaAs and GaP NWs with the ZB and WZ structures grown, respectively, along the [111] and [0001] directions. Experimentally these III-V NWs can be grown having both the ZB and the WZ structures, and also with mixed planar stackings along the growth direction, resulting in a number of different polytypes. The ZB NWs we have considered here present six $\{$101$\}$ facets on their sidewalls while the WZ NWs have six $\{$1010$\}$ facets, both wires resulting in hexagonal cross sections. The dangling bonds at the lateral surfaces were saturated by pseudo-hydrogen atoms, i.e., hydrogen atoms carrying a charge of 0.75 or 1.25 electrons, in order to mimic the $sp^3$ orbitals of the absent Ga and As(P) atoms, respectively. The diameters of the NWs were defined as those of the thinnest cylinders containing the NWs. The GaAs(GaP) ZB NWs have diameters 1.21(1.17), 1.66(1.62) and 2.12(2.06) nm, while the GaAs(GaP) WZ NWs have diameters are 1.48(1.44), 2.26(2.18), and 3.04(2.94) nm. Figure \ref{nanowires} shows representative cross sections and atomic positions along the growth directions of the largest studied NWs.

We used tetragonal supercells having the minimum size along the growth direction, i.e., $\sqrt{3}a$ for the ZB and $c$ for WZ where $a$ and $c$ are the bulk lattice parameters. In order to avoid interactions between the images at different cells, the lateral dimensions of the tetragonal cells were adjusted to accommodate a vacuum layer of approximately 10 {\AA}. Table \ref{NWs} shows the equilibrium lattice constants and the band gaps obtained for the ZB and WZ NWs. 

\begin{table}[h!]
\caption{\label{NWs} Equilibrium lattice constants and band gaps for the ZB and WZ GaAs and GaP NWs. The labels D and I stand for direct and indirect band gap, respectively. The values in parentheses in the first column refer to the diameters of the WZ NWs.}
\begin{ruledtabular}
\begin{tabular}{ccccccccc}
& \multicolumn{4}{c}{ZINC-BLENDE} & \multicolumn{4}{c}{WURTZITE} \\
  & \multicolumn{2}{c}{GaAs} & \multicolumn{2}{c}{GaP} & \multicolumn{2}{c}{GaAs} & \multicolumn{2}{c}{GaP} \\ \cline{2-3} \cline{4-5} \cline{6-7} \cline{8-9}
Diameter (nm) &  $c$({\AA})  & $E_{g}$ (eV) & $c$ ({\AA}) & $E_{g}$(eV) & $c$({\AA})  & $E_{g}$ (eV) & $c$ ({\AA}) & $E_{g}$(eV) \\  \hline
$\approx$ 1.2 (1.4) & 9.636 & 2.714(I) & 9.297 & 2.872(D) & 6.503 & 2.114(I) & 6.274 & 2.431(I) \\
$\approx$ 1.6 (2.2) & 9.665 & 2.239(I) & 9.320 & 2.357(D) & 6.509 & 1.708(I) & 6.275 & 2.023(I) \\
$\approx$ 2.0 (3.0) & 9.678 & 1.879(I) & 9.330 & 2.052(D) & 6.511 & 1.473(I) & 6.276 & 1.778(I) \\
  \end{tabular}
\end{ruledtabular}
\end{table}

\begin{figure}[h!]
\centering
\includegraphics[width=12 cm,clip]{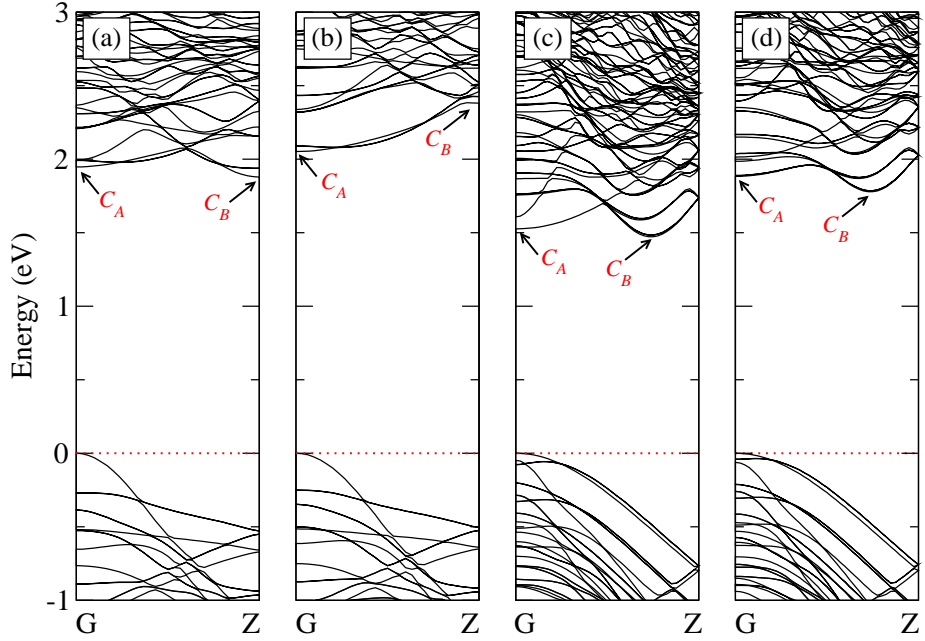}
\caption{(Color online) Calculated band structures for the largest NWs (a) ZB GaAs, (b) ZB GaP, (c) WZ GaAs and (d) WZ GaP. $C_A$ and $C_B$ indicate the two minima in the conduction band.}
\label{BandsZBWZ}
\end{figure}

\begin{figure}[h!]
\centering
\includegraphics[width=12 cm,clip] {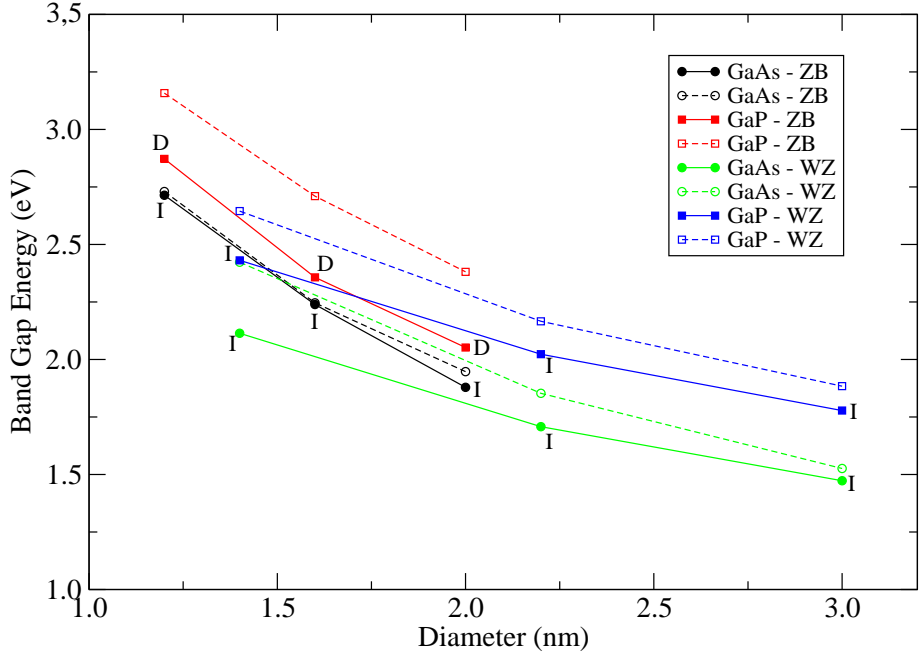}
\caption{(Color online) Band gap energy as a function of NW diameter of GaAs and GaP for zinc-blende and wurtzite structures. The labels D and I stand for direct and indirect band gap, respectively. The solid curves show the variation of $E_{C_A} -E_{VBM}$, whereas the dashed curves are the data for $E_{C_B} -E_{VBM}$. } 
\label{gapdiameter}
\end{figure}

The calculated band structures for the largest GaAs and GaP NWs with ZB and WZ structures are shown in Fig. \ref{BandsZBWZ}. The band structures for the thinner NWs are very similar and will not be shown here. As expected, the lateral quantum confinement increases the energy separation between the valence and conduction bands with respect to their bulk counterparts, increasing their band gaps. Figure \ref{gapdiameter} shows the behavior of both direct and indirect band gaps ($C_A$ and $C_B$ in Fig.\ref{BandsZBWZ}, respectively) as a function of the NW diameters. We first emphasize that, for the range of studied diameters, there is no change in the band gap character (direct or indirect) for both GaAs and GaP NWs. This allows us to restrict the analyzes hereafter to only a given diameter for each NW.

The main fact to highlight here is that the character of the band gaps in both NWs and for each structure (ZB, WZ) is opposite to those of the respective bulk counterparts. This is due to the competition between the energies of the bottom states in the two A and B valleys at the conduction band edges. 

Some of these results are in agreement with the literature. \cite{Santos,Copple,Peng, Mohammad} Santos and Schmidt \cite{Santos} investigated, by using first principles calculations, the electronic properties of ZB GaP NWs aligned along the [111] direction with diameters varying from 1.0 to 3.0 nm, approximately. They observed that for this range of NW diameters, the GaP NWs have a direct band gap in contrast to the bulk indirect band gap. The diameter dependence of the structural and electronic properties of ZB and WZ GaP NWs were also studied by Mohammad and Kat\i rc\i o\u{g}lu. \cite{Mohammad} They found, using a full potential-linearized-augmented plane waves method, a direct band gap for the ZB GaP NWs and an indirect band gap for the WZ GaP NWs. Similar studies have been performed on GaAs NWs. \cite{Copple, Peng} The authors showed that ZB [111] and WZ [0001] GaAs NWs have indirect band gap for NWs with diameters smaller than 3 nm. Further, they have demonstrated that the band structures of the NWs can be significantly tuned by applying an external strain. Indirect to direct band gap transitions can be obtained with an appropriate applied stress.

Recent theoretical studies \cite{Yang1,Yang2} showed, in contrast with our results, that ZB [111] GaP NWs have an indirect band gap, which follows the indirect band gap nature of ZB GaP bulk. For WZ [0001] GaP NWs, on the other hand, they observed an indirect to direct band gap transition when the NW diameter is approximately 2 nm.

The reasons to these differences are related to the reduced dimensionality of the NWs, which will be evidenced through (i) band folding, (ii) confinement, and (iii) surface relaxation effects. Here we will not analyze surface reconstruction effects since the NWs have their surface bonds saturated by hydrogens with fractional charges.

%%%%%%%%%%%%%%%%%%%%%%%%%%%%%%%%%%%%%%%
\subsubsection{ Contributions due to band folding effects}
%%%%%%%%%%%%%%%%%%%%%%%%%%%%%%%%%%%%%%%%

Comparing the band structures of the ZB NWs (Fig. \ref{BandsZBWZ} (a) and (b)) and ZB binary systems along the G-L direction (Figure \ref{bands_zbwz} (a) and (b)) which corresponds to the G-L [111] direction of the NWs, it is possible to see that in both cases we have a conduction valley around the G-point, much less dispersive in the case of the NWs than in the case of bulk GaAs and GaP systems, and an additional conduction valley around the L point. In GaAs the valley at G is lower in energy than the valley at L. For GaP the opposite is true. On the other hand, for the WZ NWs, the conduction edge (see Fig.   \ref{BandsZBWZ} (c) and (d)) with its two valleys, one at G and the other before Z, is completely different from the band states at the conduction edge of the WZ bulks along the G-A [0001] direction (Fig. \ref{bands_zbwz} (c) and (d)) where only one valley is present at the G point, even considering that the NWs share the hexagonal symmetry of a WZ bulk more than that of the cubic ZB bulk.

In order to understand the origin of the two valleys at the bottom of the conduction bands in the NW systems and link them to the states in the corresponding binary systems we studied the band folding upon the [111] and [0001] directions due to confinement effects in the NWs.

In the NWs there will be a discrete set of periodic motifs at directions perpendicular to the NW axis. This leads to the folding of discrete levels, parallel to the NW axis, onto to the NW band structure along its axial direction. As the NW diameter increases, more periodic motifs will appear, leading to an increasing number of folded states (that are parallel to the NW axis), along the translational symmetry direction of the NW. In the bulk limit all the bands along all the directions parallel to the NW axis will fold exactly onto the considered symmetry direction, generating the bulk band structure along that direction.

\begin{figure}[h!]
\centering
\includegraphics[height=0.85\textheight,width=0.50\textwidth,clip]{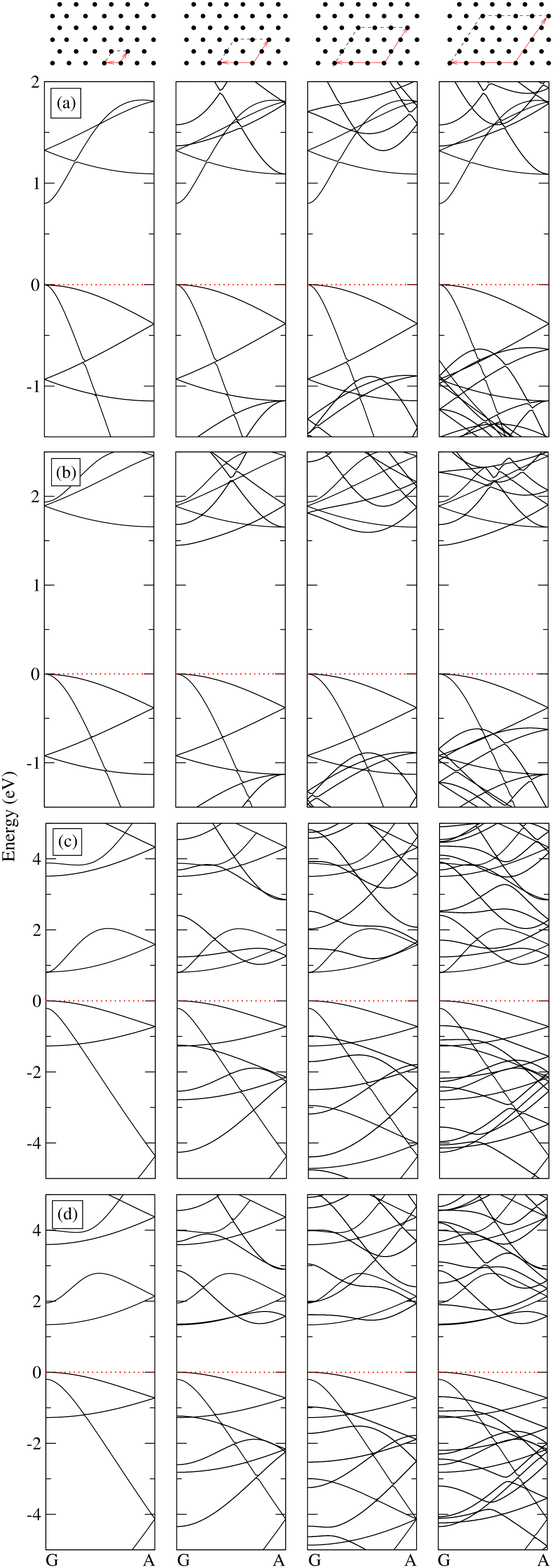}
\caption{(Color online) Calculated band structures for the unstrained (a) GaAs and (b) GaP systems in the ZB phase, and (c) GaAs and (d) GaP in the WZ phase, using a different number of atoms in the unit cells as shown on the top of the figure. The  points represent the (0001) and (111) in-plane atomic positions.}
\label{bands1}
\end{figure}

We first analyze what happens to the band structure of the binary compounds when the lattice parameters  $\bf a$ and $\bf b$ in the  planes orthogonal to the [0001] (or [111]) directions are extended. Fig. \ref{bands1} shows the band structures of unstrained GaAs and GaP in the ZB and WZ phases using a different number of atoms in the unit cells: 6(4), (24)16, (54)36 and (96)64. It can be clearly seen that multiple in-plane lattice vectors create a second valley along the G-A direction at positions similar to those of the valleys in the conduction band of the corresponding NWs. Also, interestingly, the position and shape of the valleys, near or at the A point, can be modified through different choices of the in-plane lattice vectors. Thus, we see that by extending the dimension of the unit cell in the planes orthogonal to the [0001] and [111] directions we are actually  folding onto the G-A direction the bulk bands along directions parallel to the G-A direction.

\begin{figure}[h!]
\centering
\includegraphics[height=0.85\textheight,width=0.50\textwidth,clip]{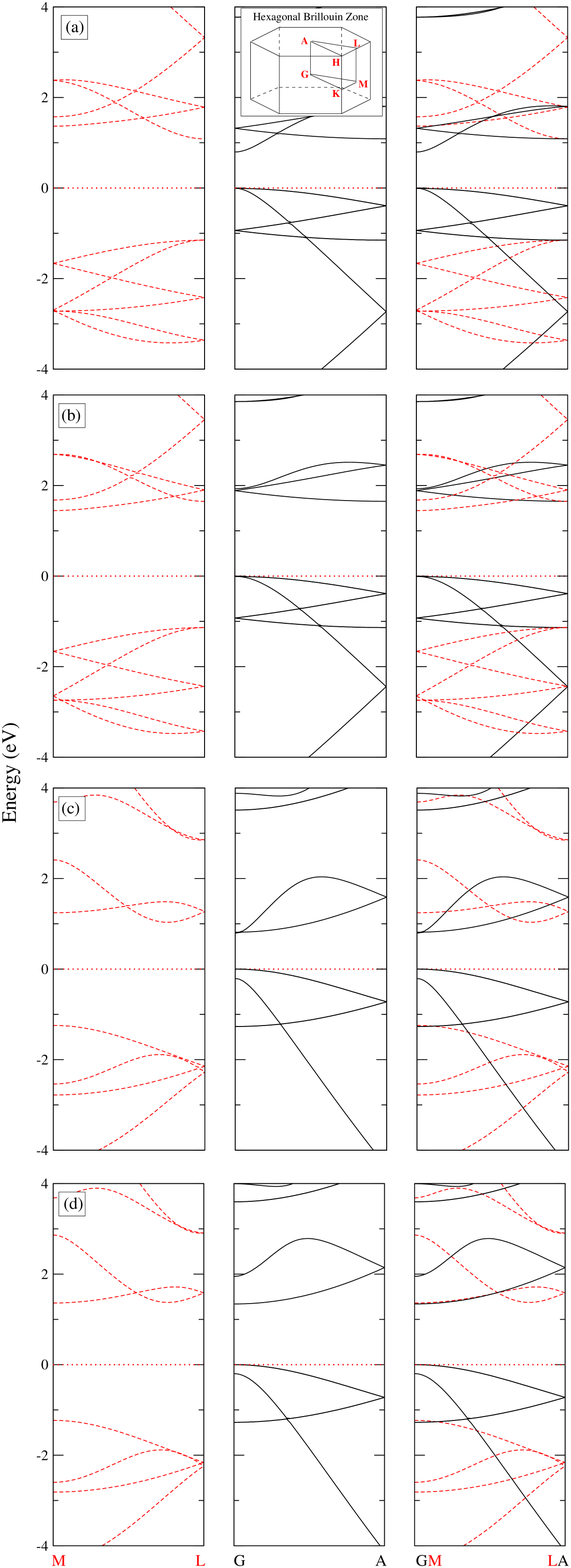}
\caption{(Color online) Calculated band structures for the unstrained: (a) ZB GaAs, (b) ZB GaP, (c) WZ GaAs, and (d) WZ GaP along the M-L direction (left),  along the G-A direction (middle), and their superposition (right). The hexagonal Brillouin zone shown in the picture indicates the M-L and G-A directions.}
\label{bands7}
\end{figure}

It is important to understand which Brillouin Zone directions fold onto the WZ [0001] and ZB [111] NW directions when we extend the in-plane lattice parameters. This may be at the origin of the $C_B$ valley at the conduction band bottom in the WZ and ZB NW band structures. To obtain this information we plot the bulk band structures relative to the G-A direction, at the center of the hexagonal Brillouin Zone, and to the M-L direction, parallel to the G-A direction but along the BZ border edge, as shown in Fig. \ref{bands7}. We can see that the resulting dispersion of the superposed band structures is the same of that obtained doubling the in-plane lattice vectors, as shown in Fig. \ref{bands1} (a) and (b).

In the case of the WZ band structures the $C_A$ valley is related to the original valley at the G point for both GaAs and GaP, and the $C_B$ valley is related to a folded band. For the ZB band structures we observe a different behavior. In the case of GaAs, the $C_A$ valley is the same G valley of the unfolded dispersion, and the $C_B$ valley is also due to the original dispersion along the G-A direction, when the ZB phase is described using the hexagonal unit cell. No contributions from the folding are apparent. In ZB GaP, instead, the $C_A$ valley derives from a folded band while the $C_B$ valley is related to the original dispersion along G-A. This shows that the direct gap at G in ZB GaP, obtained by this folding procedure, is in fact a pseudo-direct gap. We should note also that the bottom edge of the conduction bands along G-A could be lower in energy and with a slightly different shape than that estimated by simply folding only the M-L bands onto the G-A bands since not always the lowest conduction states  occur along the M-L or G-A directions but they could occur along others of the directions parallel to G-A.

%%%%%%%%%%%%%%%%%%%%%%%%%%%%%%%%%%%%%%%
\subsubsection{Contributions due to confinement and surface relaxation effects}
%%%%%%%%%%%%%%%%%%%%%%%%%%%%%%%%%%%%%%%%

We have seen that we can roughly reproduce the behavior of the valence and conduction band edges in the NWs just folding the binary bands in the appropriate way. This result can be useful since it  allows to estimate the modifications in the band structures of the NWs just using the binary band structures which are much easier to calculate and often experimentally known.  

We will now look at the contributions to the electronic structure of the GaAs and GaP NWs (ZB and WZ) arising from the spatial confinement and surface relaxations. Starting from the bulk materials, we first apply biaxial stress and look at the folded band structures. We then build NWs by cutting the biaxially stressed bulk materials. Finally, the atomic positions of the NWs are relaxed. This will allow us to analyze the electronic effects in the band structures arising from confinement and surface relaxation separately and in an unified approach for both materials.

\begin{figure}[h!]
\centering
\includegraphics[height=0.6\textheight,clip]{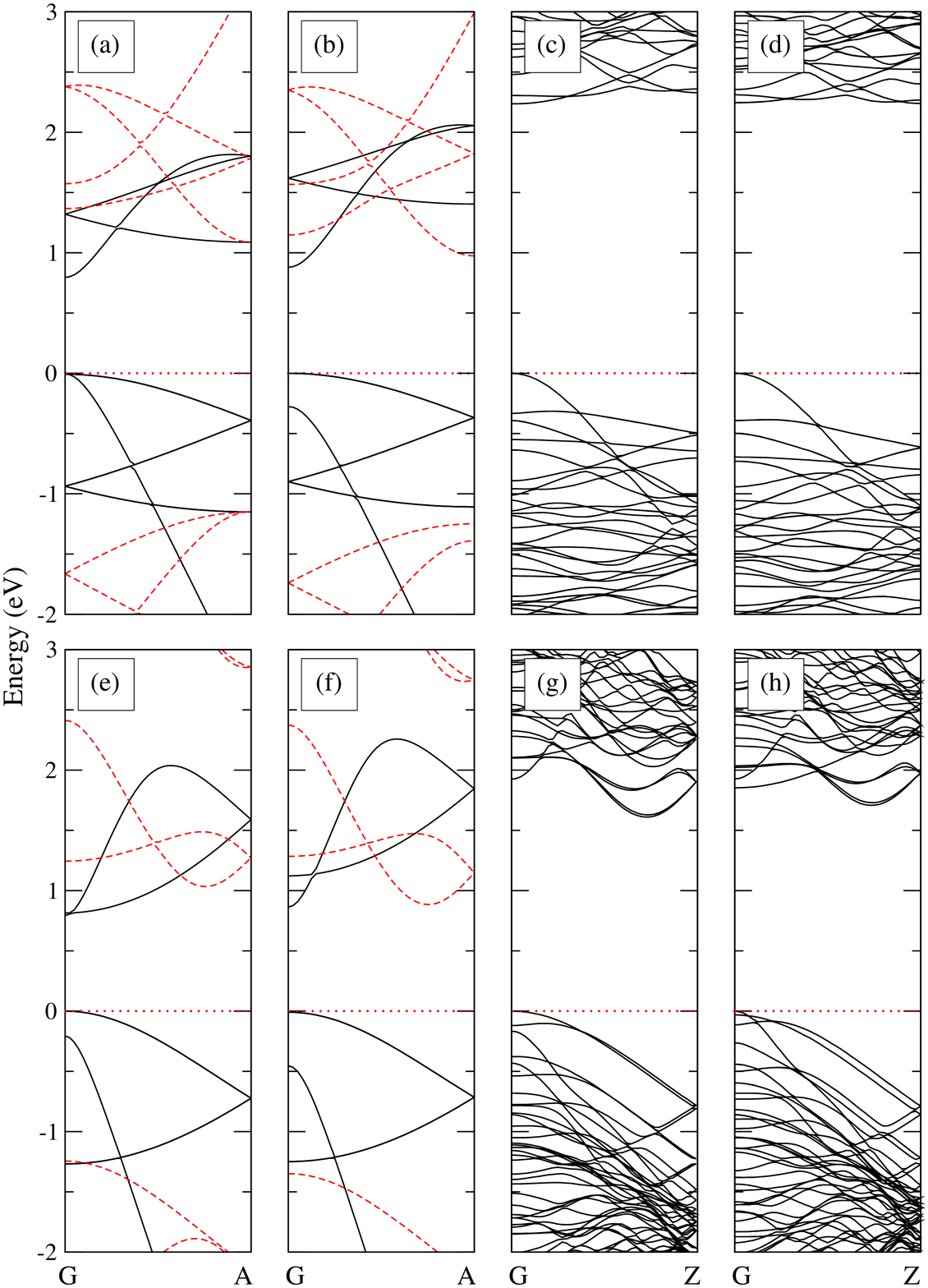}
\caption{(Color online) Band structures for the (a) ZB GaAs binary , (b) ZB GaAs binary strained to $a_{\parallel} = a_{\scriptsize{\mbox{ave}}}$, (c) ZB GaAs NW constrained to $a_{\parallel} = a_{\scriptsize{\mbox{ave}}}$, (d) ZB GaAs NW, (e) WZ GaAs binary, (f) WZ GaAs binary strained to $a_{\parallel} = a_{\scriptsize{\mbox{ave}}}$, (g) WZ GaAs NW constrained to $a_{\parallel} =  a_{\scriptsize{\mbox{ave}}}$, and (h) WZ GaAs NW. The band structures for the binaries are those obtained from an unit cell that has twice the size of the primitive cell along the direction perpendicular to that of the NW axis. }
\label{bands3}
\end{figure}

\begin{figure}[h!]
\centering
\includegraphics[height=0.6\textheight,clip]{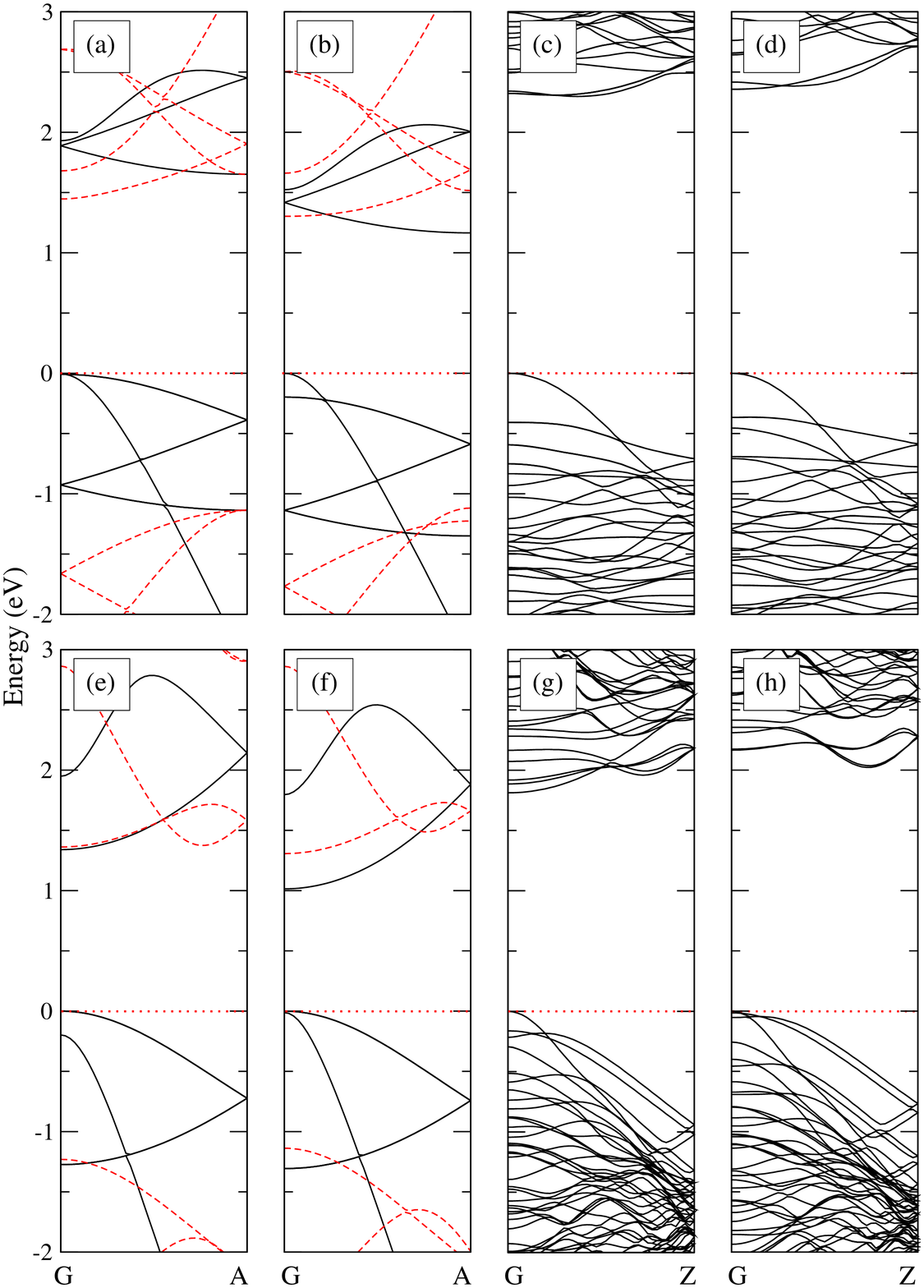}
\caption{(Color online) Band structures for the (a) ZB GaP binary , (b) ZB GaP binary strained to $a_{\parallel} = a_{\scriptsize{\mbox{ave}}}$, (c) ZB GaP NW constrained to $a_{\parallel} = a_{\scriptsize{\mbox{ave}}}$, (d) ZB GaP NW, (e) WZ GaP binary, (f) WZ GaP binary strained to $a_{\parallel} = a_{\scriptsize{\mbox{ave}}}$, (g) WZ GaP NW constrained to $a_{\parallel} =  a_{\scriptsize{\mbox{ave}}}$, and (h) WZ GaP NW. The band structures for the binaries are those obtained from an unit cell that has twice the size of the primitive cell along the direction perpendicular to that of the NW axis.}
\label{bands4}
\end{figure}

To see how the biaxial strain would change the relative energies of the $C_A$ and $C_B$ conduction valleys of the bulk materials we have calculated the folded bands for the strained GaAs and GaP binaries in both WZ and ZB phases. The results for GaAs are shown in Fig. \ref{bands3} and for GaP in Fig. \ref{bands4}. In these figures, panels (a) and (b) show the folded band structures for the ZB with their own lattice parameter and biaxially strained to $a_{\parallel} = a_{\scriptsize{\mbox{ave}}}$, respectively, while panels (e) and (f) show these same quantities for the WZ structures.


In the case of GaAs the biaxial strain is compressive in the plane orthogonal to the [111]/{0001] direction leading to an elongation of the lattice parameter along the [111]/[0001] direction. In the case of WZ GaAs the CBM at G splits with respect to the unstrained case and the split increases with the increasing of the compressive strain (not shown here). The CBM minimum at G remains the one of the original bulk dispersion. The obtained values for the energy differences at the $C_A$ and $C_B$ valleys  for ZB GaAs with $a_{\parallel} = a_{\scriptsize{\mbox{ave}}}$ relatively to the unstrained configuration are, respectively, 0.437 and 0.240 eV. For WZ GaAs, the obtained values at $C_A$ and $C_B$ with $a_{\parallel} = a_{\scriptsize{\mbox{ave}}}$ are 0.475 and 0.255 eV. 

In the case of GaP the biaxial strain is instead tensile on the planes while the lattice parameter along [111]/[0001] is shortened. In ZB GaP the minimum of the conduction band is at point A (indirect gap) and lowers with the increase of the tensile strain, while at the G point there is a swap at the CBM between a folded state (for $a_{\parallel} = a_{\scriptsize{\mbox{ave}}}$) and the original state so that the folded state becomes the lowest at G. The gap, however, remains indirect.
In WZ GaP the CBM and CBM+1 at G (original and folded band), which are almost degenerate in the unstrained binary move apart, pushing down the $C_A$ valley while the $C_B$ valley moves upward with the increasing strain.  Thus, in WZ GaP the biaxial tensile strain lowers the direct gap favoring the $C_A$ valley as CBM while in the unstrained binary the two valleys $C_A$ and $C_B$ are very close in energy. 
In the valence bands the VBM and VBM-1 at G split apart or swap under the effect of biaxial strain. These bands at the top of the valence bands belong always to the original dispersion.  

We will now look at the confinement effects on the electronic structure of the GaAs and GaP NWs starting from the strained binary systems. We have constructed  GaAs and GaP NWs, with diameters of  approximately 1.5 (2.0) nm for the ZB (WZ) structure by appropriately cutting the corresponding strained binary compounds with $a_{\parallel} = a_{\scriptsize{\mbox{ave}}}$. The dangling bonds were saturated with pseudohydrogen atoms. With these geometries we have performed three calculations: (i) one calculation where all atoms were kept fixed as in the original strained binaries, (ii) another calculation where we have allowed relaxation along the axial-direction keeping fixed the radial components of the coordinates, and (iii) one last calculation where we have also allowed the system to completely relax laterally at the side surfaces. In all cases, the  pseudohydrogen atoms were allowed to relax. These three configurations were labeled as unrelaxed, partially relaxed, and relaxed. The first calculation allows us to extract the information on how the mere lateral confinement affects the band structure of the corresponding strained binary.  The calculation (ii), where only the lattice parameter along [111] or [0001] (the $c$ dimension) is allowed to relax while the in-plane NW dimensions are kept fixed, corresponds to the biaxially strained NWs of Ref. \cite{Peng}, and enlighten the correlation between the lateral confinement and the value of the $c$ lattice parameter. The resulting band structures after relaxation along the NW axis are shown in panels (c) and (g) while the band structures for the completely relaxed NWs are shown at panels (d) and (h) of Figs. \ref{bands3} and \ref{bands4}, for the GaAs and GaP NWs, respectively.

The results of these calculations are shown in Table \ref{ZBWZrelax},  where the change of the $c$ dimension of the NWs, the band gap values and their nature (direct/indirect) are reported.  From the data in this Table it is possible to note that the lateral relaxation of the surfaces (the calculation (iii)) is the most relevant effect in the determination of the nature and width of the band gap. Indeed, for ZB GaAs NWs it allows to switch from a direct band gap (as in the unconfined corresponding strained binary GaAs) to an indirect band gap, whereas for ZB GaP NWs the band gap switches from indirect as in the corresponding strained binary to direct. In the case of WZ NWs we observe a similar behavior only in the case of GaP where the band gap remains direct when passing from the strained binary to the strained NW and changes to indirect only when the strain relaxation is allowed to take place. In the case of WZ GaAs, instead, the change in the nature of the gap (from direct to indirect) may take place already with the application of the lateral confinement from the binary to the NW and then remains indirect in the following structural relaxation steps. However, to be sure on this point we should analyze the band dispersions of the strained binary along {\it all} the parallel directions to G-A and plot the minima of the conduction band versus the k-points along G-A. The lateral relaxation leads also to an increase of the band gap more in the case of the WZ NWs than in that of ZB NWs.

\begin{table}[h!]
\caption{\label{ZBWZrelax} Equilibrium lattice constant and band gaps for the ZB and WZ GaAs and GaP NWs in three different configurations: unrelaxed, partially relaxed and relaxed. The ZB NWs have diameters $\approx$ 1.5 nm while the WZ NWs have diameters  $\approx$ 2.0 nm. The labels D and I stands for direct and indirect band gap, respectively.}
\begin{ruledtabular}
\begin{tabular}{ccccccccc}
& \multicolumn{4}{c}{ZINC-BLENDE} & \multicolumn{4}{c}{WURTZITE} \\
  & \multicolumn{2}{c}{GaAs} & \multicolumn{2}{c}{GaP} & \multicolumn{2}{c}{GaAs} & \multicolumn{2}{c}{GaP} \\ \cline{2-3} \cline{4-5} \cline{6-7} \cline{8-9}
Configuration &  $c$({\AA})  & $E_{g}$ (eV) & $c$ ({\AA}) & $E_{g}$(eV) & $c$({\AA})  & $E_{g}$ (eV) & $c$ ({\AA}) & $E_{g}$(eV) \\  \hline
Unrelaxed & 9.827 & 2.232(D) & 9.276 & 2.354(I) & 6.569 & 1.603(I) & 6.229 & 1.846(D) \\
Partial relaxed & 9.741 & 2.236(D) & 9.183 & 2.297(I) & 6.553 & 1.610(I) & 6.209 & 1.812(D) \\
Relaxed & 9.665 & 2.239(I) & 9.320 & 2.357(D) & 6.509 & 1.708(I) & 6.275 & 2.023(I) \\
\end{tabular}
\end{ruledtabular}
\end{table}

%%%%%%%%%%%%%%%%%%%%%%%%%%%%%%%%%%%%%%%
\section{Summary and Conclusions}
%%%%%%%%%%%%%%%%%%%%%%%%%%%%%%%%%%%%%%%%

In this work we have studied the electronic properties of GaAs and GaP NWs. The calculations are based on the Density Functional Theory.  The NWs sidewalls were passivated using pseudohydrogen atoms. 

We have followed the evolution of the band edge states from their dispersion in the bulk compounds to the unidimensional dispersion in the NWs. We considered two different compound semiconductors: GaAs, having a direct gap and GaP that has an indirect gap instead. Two structures corresponding to two different layer stacking along the [111]/[0001] direction and having two different symmetries (cubic and hexagonal) have been considered in this study. We have found that for both GaAs and GaP the nature of the band gap (direct/indirect) reverse in very thin NWs with respect to that of the corresponding bulks. The direct and indirect nature of the band gap is decided by the energy competition of two different valleys at the conduction band edges. These two valleys derive from the folding of the bulk bands along all the direction parallel to the [111]/[0001] direction onto the main G-A direction of the hexagonal Brillouin Zone.
Further, we have studied the effects on the valley positions due to biaxial strain, to the quantum confinement, and the atomic relaxation at the sidewalls. We found that in most cases the switch between the direct to indirect (and vice-versa) gaps are related to the surface relaxation and not to the confinement of the material. Since the surface/volume ratio decreases with the increase of the NW diameter, we expect that this effect becomes negligible in larger diameter NWs which explaines why the nature of the gap becomes the same as in the bulk for the NWs having a large diameter. This results show how important could be the engineering of the NW sidewalls in thin NWs, other than the strain, to tune the NW band edge dispersion.

\begin{acknowledgments}
Cl\'{a}udia L. dos Santos acknowledges financial support provided by  ``Conselho Nacional de Desenvolvimento Cient\'{i}fico e Tecnol\'{o}gico'' (CNPq/Brazil).
The authors wish to thank the Supercomputing Center, CINECA, Bologna, Italy, for providing computing time under the three projects IscrC-CELERON, IscrC-CONAN and IscrC-DESIRE.
\end{acknowledgments}

\end{document}